\documentclass[aip,amsmath,amssymb,reprint]{revtex4-1}
\usepackage{graphicx,isomath}
\usepackage[colorlinks=true,linkcolor=blue,citecolor=blue,urlcolor=blue,pdfborderstyle={/S/U/W 1}]{hyperref}
\usepackage[section]{placeins}
\usepackage[utf8]{inputenc}
\usepackage[T1]{fontenc}
\usepackage{mathptmx}
\usepackage{etoolbox}

\makeatletter
\def\@email#1#2{%
 \endgroup
 \patchcmd{\titleblock@produce}
  {\frontmatter@RRAPformat}
  {\frontmatter@RRAPformat{\produce@RRAP{*#1\href{mailto:#2}{#2}}}\frontmatter@RRAPformat}
  {}{}
}%
\makeatother
\begin{document}

\title{Complex localization mechanisms in networks of coupled oscillators: two case studies}
\author{Zachary G. Nicolaou$^*$}
\affiliation{Department of Applied Mathematics, University of Washington, Seattle, WA, 98195-3925, USA}
\email{zgn@uw.edu}
\author{Jason J. Bramburger}
\affiliation{Department of Mathematics and Statistics, Concordia University, Montr\'eal, QC, H3G 1M8, Canada}
\date{\today}

\begin{abstract}
Localized phenomena abound in nature and throughout the physical sciences.  Some universal mechanisms for localization have been characterized, such as in the snaking bifurcations of localized steady states in pattern-forming partial differential equations. While much of this understanding has been targeted at steady states, recent studies have noted complex dynamical localization phenomena in systems of coupled oscillators. These localized states can come in the form of symmetry-breaking chimera patterns that exhibit a coexistence of coherence and incoherence in symmetric networks of coupled oscillators and gap solitons emerging in the band gap of parametrically driven networks of oscillators. Here, we report detailed numerical continuations of localized time-periodic states in systems of coupled oscillators, while also documenting the numerous bifurcations they give way to. We find novel routes to localization involving bifurcations of heteroclinic cycles in networks of Janus oscillators and strange bifurcation diagrams resembling chaotic tangles in a parametrically driven array of coupled pendula. We highlight the important role of discrete symmetries and the symmetric branch points that emerge in symmetric models.\\[1em]
Chaos \textbf{34}, 013131 (2024) \\
\url{https://doi.org/10.1063/5.0174550}
\end{abstract}

\maketitle

\begin{quotation}
Pattern-forming mechanisms for localization give rise to important phenomena in natural and man-made systems alike. While universal snaking mechanisms for localized steady states have been well characterized, less is known about the symmetry-breaking localization leading to chimera states and oscillatory gap solitons in coupled oscillator models. In this paper, we illustrate novel bifurcation routes to localization in models of coupled oscillators, emphasizing the important role of symmetry and drawing connections with chaos theory.
\end{quotation}

\section{Introduction}

We investigate mechanisms for localization in collections of coupled ordinary differential equations. In a general setting, such models take the form 
\begin{equation}\label{CoupODE} 
	\dot x_n = F(x_n,\mu) + \varepsilon \sum_{k\in C_n} G(x_k,x_n,\mu), \quad n \in \Lambda,
\end{equation}
where $\Lambda$ is a countable index set, the set $C_n\subset\Lambda$ indicates the set of elements coupled to the element at index $n$, $\mu \in \mathbb{R}^p$ are system parameters, and $\varepsilon \geq 0$ represents the strength of coupling between connected elements. The main mechanisms typically attributed to the presence of steady states with localized features embedded in a quiescent state in systems of the form \eqref{CoupODE} derive from {\em multistability}. That is, the presence of at least two stable equilibria to $F$ at some parameter value $\mu$. In the case of coupled bistable systems over the integer lattice $\Lambda = \mathbb{Z}$, localized steady states with one \cite{bramburger2020spatially} and multiple \cite{bramburger2021isolas} regions of localization have been proven to exist. These results further fully describe the regular {\em snaking} existence curves of these solutions as one varies the parameter $\mu$,  
following similar results on localized pattern formation in partial differential equations \cite{beck2009snakes}. Related analytical results along these lines sought to investigate the effect of the choice of coupling set $C_n$ when \eqref{CoupODE} is arranged on a ring \cite{tian2021snaking} and the formation of localized square patterns of activation when $\Lambda = \mathbb{Z}^2$, \cite{bramburger2020localized}. Coupling bistable systems is a tried, tested, and true method of investigating localized steady states in systems of the form \eqref{CoupODE} which is often amenable to analytical techniques in the weakly-coupled regime $0 < \varepsilon \ll 1$ through perturbative methods.  

A significantly more difficult area of investigation is determining the mechanisms that lead to localized oscillations in coupled systems. A notable bridge between localized steady states and localized oscillations are breather solutions to the discrete nonlinear Schr\"odinger equation. In this case the gauge invariance of the Schr\"odinger equation allows one to reduce the search for time-periodic standing solitons to identifying steady states corresponding only to the amplitudes at each index \cite{shi2015existence,parker2020existence,flach1998discrete,cuevas2019discrete,parker2021standing}. However, gauge symmetry is by no means necessary to observe localized oscillation, as they have also been documented in chains of coupled mechanical oscillators \cite{clerc2018chimera,fontanela2018dark,niedergesass2021experimental,papangelo2018multiple,vakakis2001normal,shiroky2020nucleation,chiu2001synchronization,papangelo2017snaking}, where there is no known method of reducing back to the well understood steady-state case. 

From the standpoint of mathematical analysis, some simplifications can be made. First assume that the uncoupled ($\varepsilon = 0$) dynamics of \eqref{CoupODE}, given by $\dot x_n = F(x_n,\mu)$, exhibit a hyperbolic periodic solution. When considering all $n \in \Lambda$ together one arrives at a hyperbolic torus of dimension $|\Lambda|$ in \eqref{CoupODE} when $\varepsilon = 0$. Then, considering the perturbative regime $0 < \varepsilon \ll 1$ one may appeal to the theory of {\em weakly coupled oscillators} \cite{schwemmer2012theory,izhikevich2006weakly,hoppensteadt1997weakly} to understand the dynamics on the perturbed torus. Precisely, Hale's invariant manifold theorem \cite{hale2009ordinary} can be used to guarantee that the uncoupled torus persists into small $\varepsilon > 0$, meaning that the dynamics of \eqref{CoupODE} can be reduced to the phases that parametrize this torus \cite{ermentrout1991multiple,wedgwood2013phase}. The reduction to these {\em phase models} lie at the heart of the theory of coupled oscillators, with a prototypical model being the Kuramoto system \cite{kuramoto1984chemical}, and have widespread application to mathematical neuroscience \cite{hoppensteadt1997weakly,ashwin2016mathematical,bick2020understanding}.  

Complex patterns of synchrony are well-known to emerge in phase models, including those that exhibit spatial localization. Notably, symmetry breaking gives rise to interesting chimera states, which are defined as states in systems of identically coupled identical oscillators that exhibit separated regions of coexisting synchrony and asynchrony.  First characterized by Kuramoto and Battogtokh in a nonlocally coupled model\cite{kuramoto2002coexistence}, the bifurcations leading to their formation were described by Abrams and Strogatz\cite{abrams2004chimera}.  While the mathematical theory has been substantially developed in the nonlocally coupled case\cite{omel2018mathematics}, less theory has been developed for chimera states in locally coupled networks\cite{laing2015chimeras}, which give rise to interesting complex dynamics.  Recently, for example, a (large but unknown) number of localized traveling chimera states was reported to occur in a ring of Janus oscillators (a generalization of the aforementioned Kuramoto model defined later in the paper)\cite{nicolaou2019multifaceted}.  (Incidentally, traveling chimera states have also been noted in networks of nonlocally coupled oscillators \cite{xie2014multicluster,bataille2023traveling}.) In the optics literature,  on the other hand, related localized oscillatory states known as gap solitons have long attracted  attention\cite{yang2001embedded,wagenknecht2003gap,ponedel2018gap,pernet2022gap}. These states are typically created in damped and periodically driven quantum systems that exhibit band gaps and can often be associated with topological phenomena and non-Hermiticity\cite{el2018non},  but the mechanisms of formation of analogous classical states have also recently attracted interest\cite{fruchart2021non}.  Such states have been recently reported, for example, in an array of parametrically-driven pendula with periodic heterogeneity \cite{nicolaou2021heterogeneity,nicolaou2021anharmonic}. 

Here, we provide a numerical investigation of both the ring of Janus oscillators and the parametrically-driven pendulum array to document two mechanisms that can lead to spatial localization. Using numerical continuation we demonstrate that the emergence of localized pattern formation in the phase models studied herein is significantly different from the steady states that have been thoroughly examined in systems of the form \eqref{CoupODE}. Furthermore, the two cases exhibit entirely distinct routes to localization because of their differing symmetry properties.  Precisely, we find that localized traveling time-periodic patterns in the Janus oscillators come into existence through a heteroclinic bifurcation, wherein the traveling component comes from visiting neighborhoods of symmetrically related localized steady states. In the pendulum array, localized periodic solutions emerge following a secondary symmetry-breaking bifurcation out of a branch of period-doubled symmetric wave modes, and they exhibit a complex tangle of branching bifurcations leading to a myriad of attracting localized periodic states. We thus suggest that the variety of mechanisms for the localization of limit cycle solutions in coupled oscillator models may be greater than has been previously appreciated.

The remainder of this paper is organized as follows. First, in Section~\ref{continuationsec} we review the details of our numerical continuation scheme, with demonstrations of localized patterns in the Swift--Hohenberg equation. Then, in Section~\ref{sec:Janus} we define the system of Janus oscillators, discuss its symmetries, and provide our continuation results. We then do the same for the coupled pendulum array in Section~\ref{sec:Pendulum}. In Section~\ref{sec:BifDiags} conjecture what drives the complexity of the bifurcation diagrams and then we conclude in Section~\ref{sec:Discussion} with a discussion of our findings.

\section{Numerical continuation and symmetric branch points}
\label{continuationsec}
\subsection{Background theory}
Here we provide a brief introduction to numerical continuation, our primary method of analysis. These techniques are now well-established and may be already familiar to the reader. In such case, this section may be skipped and one may proceed to the main results of this paper, starting in Sec.~\ref{sec:Janus}.  

We utilize AUTO,  which implements a pseudoarclength continuation strategy \cite{doedel1991numerical}, and code producing all results in this paper is available on our GitHub repository \cite{github}. Consider first a steady-state solution $x_n^0$ to Eq.~\eqref{CoupODE}, satisfying 
\begin{equation}\label{CoupODEsteady} 
	0 = F(x_n^0,\mu_0) + \varepsilon\sum_{k\in C_n} G(x_k^0,x_n^0,\mu_0), 
\end{equation}
for an initial parameter value $\mu_0$. The implicit function theorem implies the existence of a branch of steady-state solutions $x_n^*(\mu)$ with $x_n^*(\mu_0)=x_n^0$, provided the system Jacobian matrix 
\begin{equation}
J = \begin{pmatrix} \partial F / \partial x_m  + \varepsilon \sum_k \partial G / \partial x_m 
\end{pmatrix}
\end{equation}
is nonsingular. Pseudoarclength continuation provides an efficient strategy to numerically determine such solution branches even beyond bifurcation points at which the Jacobian becomes singular.  This is achieved by parameterizing the solution branch with a new auxiliary variable $s$ (the pseudoarclength) as in $(x_n^*(s), \mu(s))$. Given a current solution $(x_n^*(s),\mu(s))$ for an initial value of $s$, the solution at $s+\delta s$ is determined by solving the extended system of equations
\begin{align}
	0 &= F(x_n^*(s+\delta s),\mu(s+\delta s)) \nonumber \\
&\quad + \varepsilon \sum_{k\in C_n} G(x_k^*(s+\delta s),x_n^*(s+\delta s),\mu(s+\delta s)), \nonumber \\
         0 &= \sum_n (x_n^*(s+\delta s)-x_n^*(s)) \delta x_n + (\mu(s+\delta s)-\mu(s)) \delta \mu  - \delta s, \label{extended}
\end{align}
via Newton's method, where $\delta s$ is the pseudo-arclength step size and $(\delta x_n, \delta\mu)$ is the ``direction vector.'' The key to the pseudo-arclength method is that the Jacobian for the extended system does not become singular at regular solution points, which include both saddle-node and Hopf bifurcation points. This is guaranteed by judicious selection of the direction vector.  Since the extended Jacobian is nonsingular, the continuation can be performed past saddle-node and Hopf bifurcations efficiently.

While only saddle-node and Hopf bifurcations emerge under one-parameter variations in generic systems, symmetric systems are capable of exhibiting other kinds of bifurcations, which correspond to branch points at which distinct solution branches coincide. Precisely, the continuation is not unique at these branch points as, in general, multiple continuation directions are possible.  Modern numerical continuation software like AUTO is capable of detecting branch points where the Jacobian of the extended system has a one-dimensional null space, which includes (codimension one) pitchfork and transcritical bifurcations. For such bifurcations, the leading order coefficient in the normal-form theory vanishes because of symmetry constraints, leading to higher-order normal forms that describe the branching solutions locally. Branch switching can usually be executed by systematically selecting alternative direction vectors at branch points.  This is achieved in AUTO by finding the roots of the determinant of the Jacobian for the extended system and solving an associated ``algebraic bifurcation equation'' for the direction vector. 

Besides such simple branch points, symmetric systems can also exhibit nonsimple branch points. One significant reason for observing nonsimple branch points in symmetric systems is that eigenvectors for symmetric solutions must generically form a representation of the corresponding symmetry group, which imposes nontrivial relationships between the eigenvalues. Thus, multiple eigenvectors can simultaneously become unstable, leading to one-dimensional unfoldings of bifurcations that would otherwise have codimension greater than one. We refer to this class of nonsimple branch points as symmetric branch points (SBPs), and we note that the problem of numerically detecting general SBPs and branch switching is still open. Equivariant bifurcation theory guarantees that there exists at least one symmetry-invariant solution branch that emerges out of such an SBP \cite{golubitsky2012singularities}. A further number of symmetry-broken solution branches also emerge out of SBPs, depending on the normal form of the bifurcation.

\subsection{Motivational example: Snaking bifurcations in the cubic-quintic Swift-Hohenberg equation}
The symmetry-breaking mechanisms for the formation of localized states can be characterized by studying the unfoldings of SBPs. To illustrate, we first briefly review the cubic-quintic Swift-Hohenberg equation
\begin{equation}
\dot{u} = ru - (1+ \partial_x^2)^2 u + 2u^3 -u^5, \label{sw}
\end{equation}
which exhibits the classic snaking bifurcation diagram of localized solutions\cite{burke2007snakes}. 
To simplify the case for the steady states of Eq.~\eqref{sw}, it is convenient to consider the continuation for the equivalent spatial ODE obtained by setting $\dot{u}=0$.  By solving for the highest order spatial derivative $u_{xxxx}$ and introducing auxiliary variables, we can express the problem in an equivalent first-order form,
\begin{align}\label{SH_ode}
u_x &= v, \\
v_x &= w, \\
w_x &= z, \\
z_x &= (r-1)u +2u^3 -u^5 -2w.
\end{align}
The uniform state $u_0=0$ is a solution for all $r$, and the Jacobian matrix for the system at $u_0$ is 
\begin{equation}
J = \begin{pmatrix} 0 & 1 & 0 & 0 \\ 0 & 0 & 1 & 0 \\ 0 & 0 & 0 & 1 \\ r-1 & 0  & -2 & 0
\end{pmatrix}. \label{SH_jac}
\end{equation}

The system is known to exhibit an SBP at $r=0$, where the eigenvalues of $J$ are $\lambda = \pm \mathrm{i}$, each with algebraic multiplicity two (a codimension two Hamiltonian-Hopf bifurcation point \cite{gaivao2011splitting}). As $r$ passes through zero, this SBP unfolds as a periodic solution branch in addition to four branches of homoclinic orbits. The homoclinic orbits of Eq.~\eqref{SH_ode} decay to zero as $x \to \pm\infty$ and exhibit recognizable snaking bifurcations as the parameter $r$ varies, leading to a multiplicity of localized steady-state solutions to the Swift-Hohenberg equation.

We note that Eq.~\eqref{sw} is invariant under reversals in the spatial coordinate $(x',u')=\pi_1(x,u)=(-x,u)$ and in the dependent variable  $(x',u')=\pi_2(x,u)=(x,-u)$ (as well as the composed reversal $(x',u')=\pi_3(x,u)=(-x,-u)$).   The invariance of the fixed point $u(x)=0$ under spatial reversals, in particular, has special implications for the structure of the eigenvalues of Eq.~\eqref{SH_jac}. The eigenvectors of $J$ must form a representation for the symmetry group and the eigenvalues must be mapped to each other under a compatible group action.  For the $\pi_1$ and $\pi_3$ symmetries, each eigenvalue $\lambda$ is mapped to an eigenvalue $-\lambda$. Since the system is real,  complex conjugation also maps eigenvectors to other eigenvectors corresponding to conjugate eigenvalues. 

Consequently,  $J$ in Eq.~\eqref{SH_jac} can exhibit either i) four non-real and non-imaginary complex eigenvalues,   ii) four imaginary (neutrally stable) eigenvalues,  iii) four real eigenvalues,  or iv) two imaginary and two real eigenvalues. (It cannot exhibit two imaginary eigenvalues or two real eigenvalues coexisting with two non-real and non-imaginary eigenvalues.) Thus, whenever a pair of complex conjugate eigenvalues approaches the imaginary axis in a Hopf-like bifurcation, a pair of sign-reversed eigenvalues simultaneously approaches the imaginary axis at the same point,  and the representation can change from type (i) to type (ii) at a SBP. Since the equation depends on the spatial derivatives linearly, the invariance under spatial reversals is sufficient to guarantee an Hamiltonian form, leading to the particular Hamiltonian-Hopf bifurcation noted above. More generally, symmetry considerations can restrict the possible bifurcations that a system can exhibit, but symmetry alone does not guarantee that such bifurcations will occur.

\begin{figure*}[hbt]
\includegraphics[width=1.9\columnwidth]{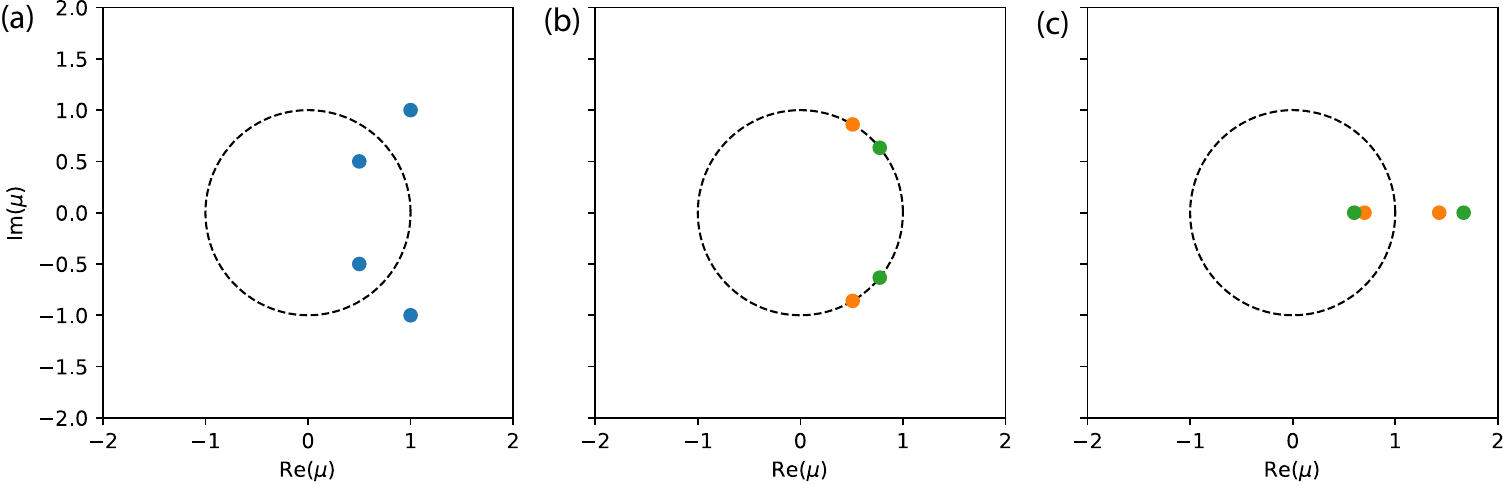}
\caption{Symmetry constraints on Floquet multipliers for time reversal invariant limit cycles.  Multipliers must appear in i) groups of four complex multipliers (shown in a) that map to each other under conjugations and inversions, ii) groups of two complex multipliers (shown in b) with unit norm that map to each other under conjugation and inversion,  iii) groups of two real Floquet multipliers (shown in c) that map to each other under inversion and to themselves under conjugation, or iv) groups of single unit multipliers  (not shown) mapping to themselves under inversions and conjugations. \label{fig1}}
\end{figure*}

\subsection{Limit-cycle stability,  continuation, and invariance}
In the remainder of the paper, we focus on the emergence of localized limit cycles in coupled oscillator models described by Eq.~\eqref{CoupODE} rather than steady states.   The stability of limit-cycle solutions to Eq.~\eqref{CoupODE}, i.e. solutions with $x_n^0(t+T)=x_n^0(t)$ with period $T$,   is determined by their Floquet multipliers.  Take an arbitrary starting point on the cycle $x_n^0(0)$ at $t=0$ and consider the trajectory $x_n(t)$ following a small perturbation off of the limit cycle with initial point $x_n(0) = x_n^0(0) + \delta_n(0)$. We then consider the linearized flow map $\Phi(t)$ determining the evolution of the perturbation after a time $t$, such that $\delta_n(t) \equiv x_n(t)-x_n^0(t) = \sum_m \Phi_{nm}(t) \delta_m(0) $. The Floquet multipliers and eigenvectors are the eigenvalues and eigenvectors of the circuit map $\Phi(T)$ corresponding to the flow map after one period.  

The numerical continuation of limit cycles requires small generalizations of the pseudoarclength method described above, which are implemented well in AUTO. The limit cycle is discretized using a (fourth order) collocation method, and an additional degree of freedom is introduced to allow the period $T$ of the limit cycle to vary during the continuation. Furthermore, in addition to temporal periodic boundary conditions, a phase-fixing integral condition is introduced to resolve the invariance under time translations, 
\begin{equation}
\int_0^T \sum_n \dot{x}_n^{\mathrm{old}}(t)(x_n(t)-x_n^{\mathrm{old}}(t))dt = 0,
\end{equation}
where $x_n(t)$ is the sought limit-cycle solution, $x_n^{\mathrm{old}}$ is the solution from the previous step, and $\dot{x}_n^{\mathrm{old}}$ is time derivative of the previous solution. This condition minimizes the $L^2$ distance between ${x}_n$ and ${x}_n^{\mathrm{old}}$ over time shifts \cite{doedel1981auto,doedel1991numerical2}. 

We next briefly review more formal notions of symmetry transformations, symmetry-invariant solutions, and implications to stability analysis, mainly following the framework of Ref.~\cite{olver1993applications}.  We regard any transformations on the combined space of independent and dependent variables as a symmetry transformation if it maps any solution of the given set of differential equations to another solution. If a set of differential equations admits a symmetry, it is generally possible to seek symmetry-invariant solutions as well. In particular, the compatibility of the conditions that a solution satisfies both the original set of differential equations and the constraints of being mapped to itself under a transformation is guaranteed by the symmetry condition (this procedure of reducing equations based on symmetry invariance can be formalized with quotient spaces).  

For limit cycle solutions $x_n^0(t)$, invariance under a symmetry map $\pi$ means that each point on the limit cycle is mapped to another point on the limit cycle. In particular, the initial point is mapped to some time-shifted point,  $\pi(x_n^0(0)) =  x_n^0(\tau)$.Thus, given a Floquet eigenvector $v$ corresponding to the base point at $t=0$, the pushforward under the symmetry map $d\pi$ will be a Floquet eigenvector at the base point at $t=\tau$. The pushforward of the reverse flow map $d\Phi(-\tau)$ will then map the Floquet eigenvector back to the initial base point $v' = d\Phi(-\tau) [d\pi [v]]$,  and the Floquet eigenvectors will thus form a representation of the symmetry group, with each eigenvector mapping to another eigenvector or itself under each symmetry map. 
\begin{figure*}[hbt]
\includegraphics[width=1.9\columnwidth]{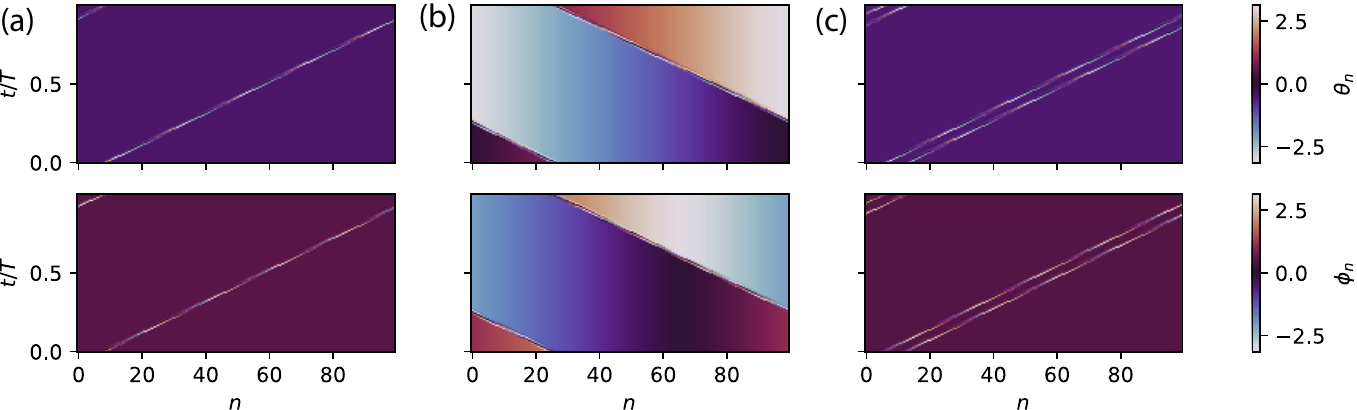}
\caption{Traveling chimeras in a ring of $N=100$ Janus oscillators with $\sigma=0.3$. The state in (a) travels to the right, exhibiting no pitch and a single defect; the state in (b) travels to the left, exhibiting a pitch and a single defect; and the state in (c) travels to the right, exhibiting no pitch with two nearby defects. (Multimedia view) Animation of the evolving phases for each solution.  \label{fig_janus}}
\end{figure*}

The Floquet multipliers corresponding to $v$ and $v'$ must also be compatible under the symmetry map. For example, we will see in the case of Janus oscillators that the model exhibits time-reversal symmetries. In this case, the circuit map $\Phi(T)$ is mapped to its inverse $\Phi(-T)$, and the multiplier $\mu$ corresponding to eigenvector $v$ will thus be mapped to a multiplier $1/\mu$ corresponding to eigenvector $v'$.  As illustrated in Fig.~\ref{fig1}, there is then an exact analogy with the reversal symmetry noted in the spatial problem of the Swift-Hohenberg equation, with the Floquet multipliers corresponding to exponentiated eigenvalues $\mu=e^{\lambda T}$ appearing as groups of four complex multipliers, groups of two multipliers constrained to the unit circle, or groups of two multipliers constrained to the real line.  In Sec.~\ref{sec:Janus} below,  we find interesting examples in the ring of Janus oscillators where all the Floquet multipliers are constrained to lie on the unit circle, and thus the invariant limit cycle is maximally neutrally stable.

\subsection{Modifications to AUTO}
The Floquet multipliers for limit cycles are evaluated efficiently in AUTO through back substitution while solving a linear system involving the discretized Jacobian\cite{fairgrieve1991ok}. Local bifurcations occur when Floquet multipliers cross the unit circle in the complex plane. There is always a trivial Floquet multiplier at 1 corresponding to translations along the phase of the cycle, and AUTO can detect saddle-node, torus, period-doubling bifurcations, and simple branch points as above. The simple branch points are identified by determining the zeros of the determinant of the extended Jacobian using a bracketed Mueller method when a sign change occurs, which works well in many cases. However, we find that the sign of the determinant in our models below becomes numerically unstable, giving rise to the identification of spurious branch points. On the other hand, the Floquet multipliers near the unit circle are numerically stable\cite{fairgrieve1991ok}, so it is reasonable to monitor the Floquet multiplier directly. This requires small changes to the AUTO source code in order to recompute the multipliers between the steps in the Mueller method identifying zeros of user-supplied special functions, as detailed in our GitHub repository \cite{github}. These modifications also enable us to detect additional SBPs in which two real multipliers simultaneously cross the unit circle. Such SBPs were previously undetected in AUTO since the determinant of the Jacobian does not change signs when two real eigenvalues simultaneously change signs. We mark such SBPs with $\times$ symbols in the bifurcation diagrams below. 

Since the computations we perform in this work are expensive,  we limit each continuation to $10$ limit points and to $10$ branch points in the ring of Janus oscillators and $20$ limit points and to $20$ branch points in the coupled pendulum array (which is comparatively less expensive). We also attempt to perform branch switching at each simple branch point, and we keep track of visited bifurcation points, terminating the continuation if a point is revisited. Finally, we terminate the continuation if the period of the cycles grows too large ($T>1500$ for the ring of Janus oscillators here). 

\section{Janus oscillators on a ring}\label{sec:Janus}

\subsection{Previous observations}
The Janus oscillator model was recently introduced\cite{nicolaou2019multifaceted} as a minimal model to investigate the implications of antiferromagnetic order in synchronization. Antiferromagnetic order refers to the order observed in many-body systems that naturally arrange into a configuration with alternating orientations, as seen in antiferromagnetic spin systems. The synchronization dynamics in systems with antiferromagnetic order have been proposed to have implications for spin waves in antiferromagnets and other condensed matter systems as well as for biological systems with two kinds of agents that form pairs and act together as a unit in collective dynamics.  The basic unit in the Janus oscillator model is a pair of phase oscillators $\theta_i$ and $\phi_i$ with differing natural frequencies $\omega_1$ and $\omega_2$ which are sine-coupled with an internal coupling constant $\beta$. When coupling Janus oscillators together, we include a sine-coupling between oscillators of differing types to impose the antiferromagnetic order. The simplest scenario is thus a one-dimensional ring of Janus oscillators, defined by the model
\begin{align}
\dot{\theta}_n &= \omega_1 + \beta\sin(\phi_n - \theta_n) + \sigma \sin(\phi_{n+1}-\theta_n), \label{janus1}\\
\dot{\phi}_n &= \omega_2 + \beta\sin(\theta_n - \phi_n) + \sigma \sin(\theta_{n-1}-\phi_n), \label{janus2}
\end{align}
with $n=N+m$ identified with $n=m$, describing periodic boundary conditions. For simplicity here, we equate the internal and external coupling strengths to a single coupling strength $\beta=\sigma$.  We impose this constraint only to reduce the number of free parameters for this study; qualitatively similar results hold if we, say, fix $\beta$ and allow only $\sigma$ to vary as done in Ref. \cite{nicolaou2019multifaceted}. Furthermore, by entering a rotating frame and rescaling the time and coupling constant, we can take the frequencies as $\omega_1=1/2$ and $\omega_2=-1/2$.  Previous investigations of Eq.~\eqref{janus1}-\eqref{janus2} have shown that random initial conditions relax to a plethora of attracting solutions for intermediate coupling strengths. Three such solutions for a ring with $N=100$ Janus oscillators with $\sigma=0.3$ are shown in Fig.~\ref{fig_janus}(Multimedia view), along with animations of the phases for each solution.
These solutions consist of a large group of synchronized (phase-locked) oscillators that coexist with a highly localized and traveling group of oscillators that do not synchronize with the others, thus forming a traveling chimera state. The various solutions differ based on the amount of phase twisting pitch in the synchronized groups, the number and relative positions of asynchronous groups, and the dynamics within the asyncronous groups.  Many differing configurations of synchronous and asynchronous groups are observed to be attracting, and when quasistatically varying $\sigma$, these attracting solutions continue over a range of $\sigma$.

\subsection{Continuation equations}
To regularize the $2\pi$-discontinuity corresponding to rotations in the phase equations, we employ a complex representation $z_n=e^{i\theta_n}$ and $w_n=e^{i\phi_n}$ for numerical continuation. We then consider complex equations, taking the form of \eqref{CoupODE},
\begin{align}
\dot z_n &= z_n\left( \frac{\mathrm{i}}{2} + \frac{\sigma}{2}\left(w_nz_n^*-z_nw_n^* + w_{n+1}z_n^*-z_nw_{n+1}^*\right)\right)  \nonumber\\
&\quad + \gamma\left(1-z_nz_n^*\right)z_n, \label{eom1} \\
\dot w_n &= w_n\left( \frac{-\mathrm{i}}{2} + \frac{\sigma}{2}\left(z_nw_n^*-w_nz_n^* + z_{n-1}w_n^*-w_nz_{n-1}^*\right)\right) \nonumber \\
&\quad + \gamma\left(1-w_nw_n^*\right)w_n, \label{eom2}
\end{align}
Denoting the polar coordinates as $z_n = \rho_n{\mathrm e}^{\mathrm{i}\theta_n}$ and $w_n = \eta_n{\mathrm e}^{\mathrm{i}\phi_n}$, a straightforward change of variables leads to the polar equations of motion for the amplitudes $\dot \rho_n = \gamma \rho_n \left(1-\rho_n^2\right)$ and $\dot \eta_n =  \gamma \eta_n \left(1-\eta_n^2\right)$. Note that the amplitude dynamics decouples from the phases and are attracted to the fixed points $\rho_n=1$ and $\eta_n=1$ (we fix $\gamma=1$ in numerics).  Likewise, the phase dynamics reduce to Eqs.~\eqref{janus1}-\eqref{janus2}. 

The ring of Janus oscillators possesses a rich group of discrete symmetries. First, the ring is invariant under the obvious rotational symmetries $(\theta'_n,\phi'_n,t')=\pi_R(\theta_n,\phi_n,t) \equiv (\theta_{n+1},\phi_{n+1},t)$, taking the periodic boundary conditions into account with $\theta_{N}\equiv\theta_0$ and $\phi_N\equiv\phi_0$. The ring is also invariant under the time/parity reversal $\pi_1$ given by $(\theta_n',\phi_n',t') = \pi_1(\theta_n,\phi_n,t) \equiv (\pi+\phi_{N-n},\theta_{N-n},-t)$. Since this map reverses the direction of time, stable solutions are mapped to unstable solutions under $\pi_1$.  The ring is also invariant under the parity/sign reversal $(\theta_n',\phi_n',t') = \pi_2(\theta_n,\phi_n,t) \equiv (-\phi_{N-n},-\theta_{N-n},t)$. Since the direction of time is preserved by this map, the map $\pi_2$ takes stable solutions to other stable solutions and reverses the direction of the attracting traveling chimera states.  We employ the complex Kuramoto order parameter
\begin{equation}\label{OrderParam}
    re^{i\Theta}\equiv \frac{1}{2N} \sum_n \left(e^{\mathrm{i}\theta_n}+e^{\mathrm{i}\phi_n}\right)
\end{equation} 
as a state space projection to visualize solution branches. The magnitude $r$ is large when all the phases in the system are similar, and small when the phases are more uniformly distributed, so the time-averaged value of $r$ can be regarded as a measure of the overall synchrony of a periodic solution.  Note that the parity/sign reversal symmetry leaves $r$ invariant, so there are two branches of solutions corresponding to each line in the bifurcation diagrams below.  Lastly, there is a second parity reversal $(\theta_n',\phi_n',t') = \pi_3(\theta_n,\phi_n,t) \equiv (\theta_{N-n+1},\phi_{N-n},t)$, which exchanges the roles of the two coupling terms (this symmetry is only present when the internal and external coupling constants are identical). Composition of the various reversal symmetries leads to a total of seven reversal symmetries which, with the identity element, form the group $\mathbb{Z}_2\times \mathbb{Z}_2 \times \mathbb{Z}_2$, up to rotations.

Chimera solutions that are invariant under any of these symmetries play a special role in the system, as we detail below. Since the traveling chimera state solutions possess a direction of travel, they can only be invariant under those symmetries that simultaneously reverse time and parity or those that reverse neither. There are three such symmetries, given by $\pi_1$, $\pi_4\equiv\pi_1\circ \pi_2 \circ \pi_3$ with $\pi_4(\theta_n,\phi_n,t)=(\pi-\theta_{N-n+1},-\phi_{N-n},-t)$, and $\pi_5\equiv\pi_2\circ \pi_3$ with $\pi_5=(-\phi_n,-\theta_{n+1},t)$. 

Since the phase equations depend only on phase differences, the equations are also invariant under continuous global phase rotations $\theta_i\to\theta_i+\psi$ and $\phi_i\to \phi_i+\psi$.  This means that limit cycle attractors and fixed point attractors will have a neutrally stable perturbation direction corresponding to the global phase rotation. For limit cycle attractors, with the additional neutral perturbation corresponding to shifting the phase of the cycle itself $\theta_i\to\theta_i+\epsilon\dot{\theta_i}$ and $\phi_i\to\phi_i+\epsilon\dot{\phi_i}$, there will be two unit Floquet multipliers for all parameter values, rendering all points singular in the pseudo-arclength continuation method employed by AUTO (correspondingly, the total phase $\sum_n \left(\theta_n + \phi_n\right)$ is a conserved quantity). We can remove this degeneracy by moving into a reference frame that rotates at the speed of oscillator $z_0$. Define quantities $\tilde{z}_n = z_n/z_0$ and $\tilde{w}_n = w_n/z_0$ (whose phases are, respectively, $\theta_n-\theta_0$ and $\phi_n-\theta_0$).  Then $\dot {\tilde z}_n = \dot z_n / z_0 - \left(z_n/z_0\right)\left( \dot z_0/z_0\right)$ and $\dot {\tilde w}_n = \dot w_n / z_0 - \left(w_n/z_0\right) \left(\dot z_0/z_0\right)$. Assuming, without loss of generality, that $z_0$ is initialized with $z_0=1$, the $4N-2$ real Cartesian coordinates $\tilde{z}_n = x_n+iy_n$ and $\tilde{w}_n=u_n+iv_n$ evolve according to 
\begin{align}
\dot{x}_n&=-y_n\left(\sigma(x_n(v_n+v_{n+1})-y_n(u_n+u_{n+1})-v_{0}-v_{1})\right) \nonumber \\
&\quad +\gamma(1-x_n^2-y_n^2)x_n,  \label{janus_cont1}\\
\dot{y}_n&=x_n\left(\sigma(x_n(v_n+v_{n+1})-y_n(u_n+u_{n+1})-v_{0}-v_{1})\right) \nonumber \\
&\quad +\gamma(1-x_n^2-y_n^2)y_n, \\
\dot{u}_n&=-v_n\left(-1+\sigma(u_n(y_n+y_{n-1})-v_n(x_n+x_{n-1})-v_{0}-v_1)\right)\nonumber \\
&\quad +\gamma(1-u_n^2-v_n^2)u_n, \\
\dot{v}_n&=u_n\left(-1+\sigma(u_n(y_n+y_{n-1})-v_n(x_n+x_{n-1})-v_{0}-v_1)\right)\nonumber \\
&\quad +\gamma(1-u_n^2-v_n^2)v_n. \label{janus_cont4}
\end{align}
Note the limit cycle solutions of Eqs~\eqref{janus_cont1}-\eqref{janus_cont4} only exhibit periodicity in the phase differences between the oscillators, not necessarily in the absolute phases $\theta_n$ and $\phi_n$. For example, the initial and final states in the limit cycle shown in Fig.~\ref{fig_janus}(b) differ by a uniform phase shift of $\pi$.

\begin{figure*}[t!]
\includegraphics[width=1.5\columnwidth]{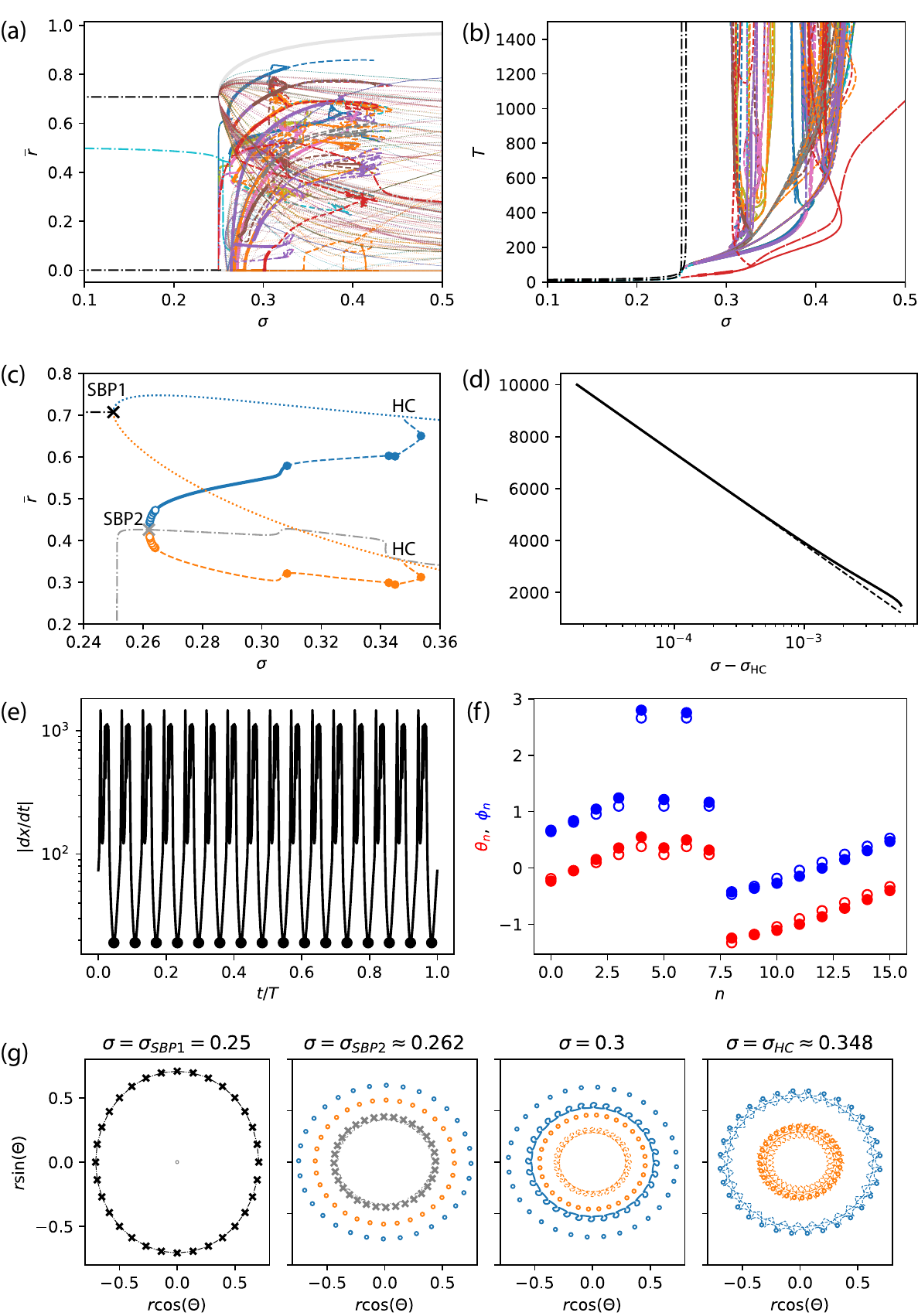}
\caption{Bifurcation diagram for the ring of Janus oscillators. (a-b) Time-averaged Kuramoto order parameter $\overline{r}$ from Eq.~\eqref{OrderParam} (a) and period $T$ (b) for all identified solution branches. Thick lines show stable limit cycles, dashed lines show unstable limit cycles, dash-dotted lines show neutrally stable invariant limit cycles, and thin dotted lines show unstable steady states. (c) Example solution branch as in (a), with solid circles denoting fold bifurcation points and simple branch (transcritical and pitchfork) bifurcation points, open circles denoting torus bifurcation points, and $\times$ symbols showing SBPs.  (d) Logarithmic fit for the period $T$ vs the distance $\sigma-\sigma_{HC}$ to the point HC in (c). (e) Rate of change norm vs. time for the large-period limit cycle of the branch in (c), with $T=1500$ and $\sigma\approx \sigma_{HC}\approx0.3480$.(f) Numerically identified steady state (open circles) starting from the minima identified in (e) (filled circles), which become connected by heteroclinic orbits at the heteroclinic bifurcation point $\sigma=\sigma_{HC}$. (g) Complex Kuramoto order parameter projection of the limit cycles and fixed points undergoing bifurcations in the exemplary solution branch, with dash-dotted, solid, and dashed lines showing invariant, stable, and unstable limit cycles, respectively, open circles showing unstable steady states,  $\times$ symbols showing SBPs, and the colors corresponding to (c). \label{fig2}}
\end{figure*}
\subsection{Continuation results}
To generate initial limit cycles, we simulate $100000$ random initial conditions with $\sigma=0.3$ for a period of $25000$ time units.  Because numerical continuation for limit cycles is expensive, we must use a relatively small value of $N=16$ here, but we suggest that the essential features of the localized states like that in Fig.~\ref{fig_janus} will not be significantly perturbed.   We numerically identify the period of any resulting limit cycles, and we evaluate the complex order parameters 
\begin{equation}
r_\pm\equiv\left(\sum_n e^{\mathrm{i}(\theta_n\pm2\pi n/N)}+e^{\mathrm{i}(\phi_n\pm2\pi n/N)}\right)/2N,
\end{equation}
in addition to the complex Kuramoto order parameter. We then evaluate the time-averaged norms and the number of $2\pi$ windings of the order parameters over a period of the cycles, and we bin final states according to these solution measures to select the $64$ top attracting chimera states in the sample. These solution measures successfully distinguish between the symmetry-related chimera states, and we observe that states come in pairs or sets of four related by the various parity-reversing symmetries.
These states are used as starting points for continuation with respect to $\sigma$ in AUTO. Figures \ref{fig2}(a)-(b) show the order parameter and the period for the solution branches corresponding to these initial limit cycle solutions and their subsequent branch-switching branches, with thick lines indicating stable states. The bifurcation diagram is quite complex, with many solution branches undergoing many bifurcations. 

\subsubsection{Exemplary solution branch}
\begin{figure*}[htb]
\includegraphics[width=1.9\columnwidth]{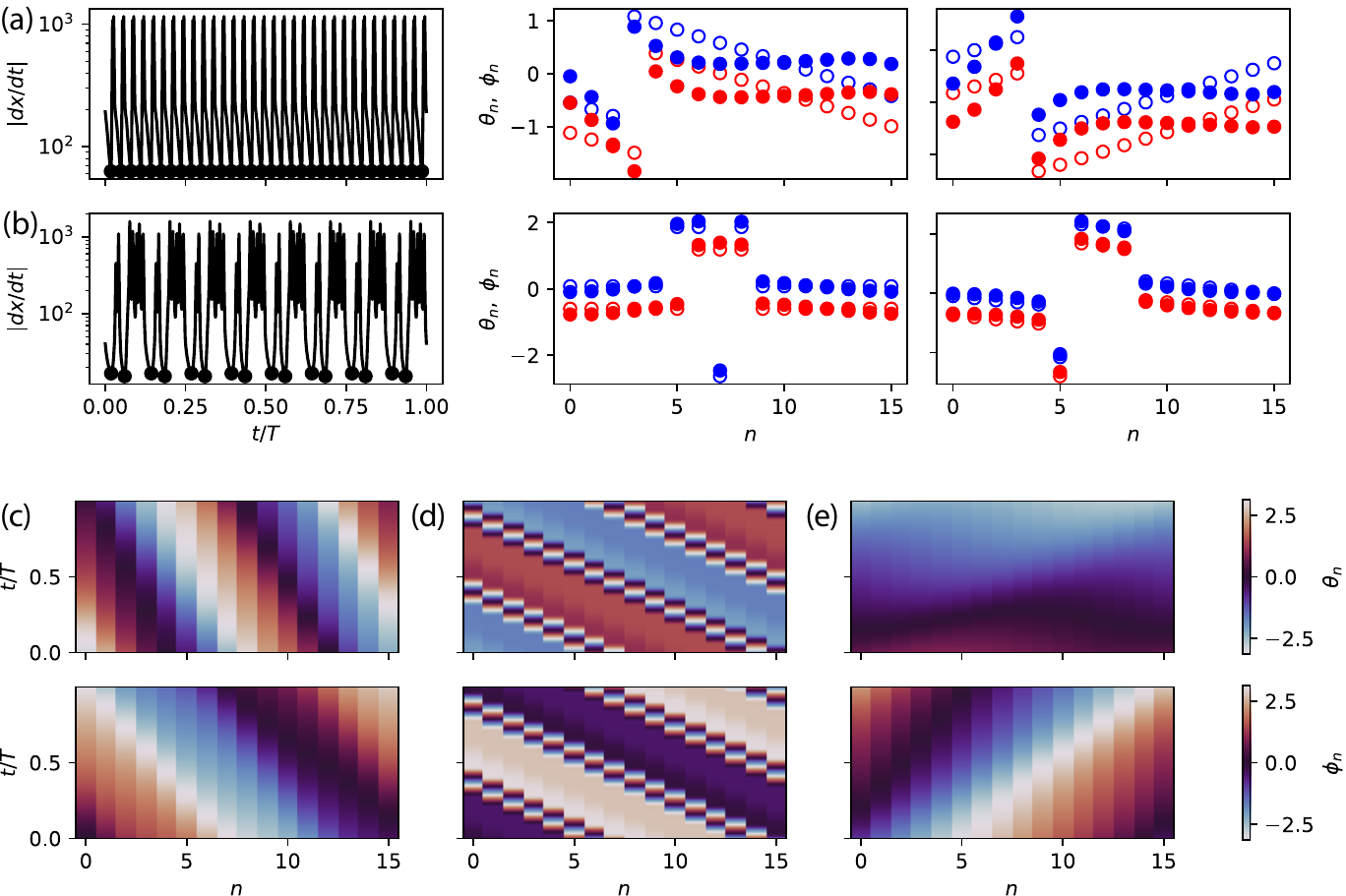}
\caption{(a-b) Steady states identified in heteroclinic bifurcations for other solution branches, as in Fig.~\ref{fig2}(e-f). The top attracting chimera state (a) at $\sigma\approx\sigma_{HC2}\approx 0.4237$ alternatingly visits reflections of a twisted state with a single defect, while another branch of unstable chimera states at  $\sigma\approx\sigma_{HC3}\approx 0.3986$ (b) alternatingly visits entirely distinct unstable steady states.  (c-e) Neutrally stable invariant chimera states identified from continuation, as in Fig.~\ref{fig_janus}.  The traveling wave invariant chimera (c) at $\sigma=0.1$ follows the second SBP of the exemplary solution branch; a cluster-twisted invariant chimera (d) at $\sigma=0.2644$ branches into several stable chimera states for larger $\sigma$; and a second traveling wave invariant chimera state (e) is prominent at $\sigma=0.1$.  \label{janus_invariants}}
\end{figure*}

One exemplary solution branch is shown in Fig.~\ref{fig2}(c). For this branch, the limit cycle loses stability with decreasing $\sigma$ and turns around at a limit point, which is an SBP. This SBP is a consequence of the time-reversal symmetries $\pi_4$, which maps the stable limit cycle to an unstable twin. A $\pi_4$-invariant chimera solution emerges out of the SBP, which is neutrally stable and not attracting. The invariant chimera continues to a second SBP that occurs near $\sigma=0.2512$ and $r=0.0134$.  This second SBP corresponds to an unusual period tripling of another invariant solution branch, which is a traveling wave solution with differing twists in the $\theta_n$ and $\phi_n$ variables. This invariant traveling wave can be continued back to $\sigma=0$ (see Fig.~\ref{janus_invariants}(c) below). Such nongeneric bifurcations are possible on invariant solution branches when the Floquet eigenvectors form certain representations of the time-reversing symmetry groups. Since $\pi_4$ will map a Floquet multiplier to its inverse by virtue of its time inversion,  $\pi_4$-invariant Floquet subspaces can be constrained to lie on the unit circle, and are thus able to pass through the otherwise nongeneric point $e^{2\pi \mathrm{i}/3}$ as $\sigma$ varies, leading to period tripling. Incidentally, we also confirmed that a period-quadrupled branch can be continued from analogous SBPs where the Floquet multipliers pass through $e^{2\pi \mathrm{i}/4}$. 

Returning to the initial stable limit cycles and now increasing $\sigma$, several simple branch points occur while the period $T$ of the cycle increases quickly, making further continuation difficult. To overcome this challenge, we periodically increase the number of mesh points proportional to the period and continue up to $T=10000$.   The rate of divergence in the period appears to be logarithmic, as shown by the fit in Fig.~\ref{fig2}(d) and suggesting the presence of a global bifurcation. To investigate the source of this phenomenon in more detail, we define a measure of the rate of change of the solutions 
\begin{equation}
\lvert {d x}/{ dt}  \rvert {\equiv} \sqrt{\sum_n  \left(\dot{\theta}_n^2+\dot{\phi}_n^2 \right) }. 
\end{equation}
Figure \ref{fig2}(e) shows the evolution of $\lvert dx/dt\rvert$ over a period of the limit cycles for $\sigma=0.3480$.  We find that the limit cycle solution periodically slows down significantly during its evolution, marked by minima in $\lvert dx/dt \rvert$ that are several orders of magnitude smaller than the average value.  By employing root finding starting at these minima, we can successfully identify steady-state solutions, as shown in Fig.~\ref{fig2}(f). The sequence of steady states visited in the limit cycle corresponds to rotations $\pi_R$ of each other composed with global phase rotations. Thus, we posit that a heteroclinic cycle exists between the rotated steady-state solutions for sufficiently large $\sigma$ and the limit cycles emerge via a heteroclinic global bifurcation with decreasing $\sigma$ near $\sigma_{HC}\approx 0.3480$ for the exemplary solution branch. 

Continuing the steady states involved in the heteroclinic cycles back to smaller $\sigma$, we find that many branches coalesce at another SBP at $\sigma=0.25$, marked with the black $\times$ in Fig.~\ref{fig2}(c). This SBP occurs where the stable and unstable twin of synchronized steady-state solutions coincide. The synchronized solutions $\theta_n=\theta$ and $\phi_n=\phi$ satisfy  $\dot{\theta}-\dot{\phi}=1+4\sigma\sin(\theta-\phi)$. For $\sigma<0.25$ the two groups of oscillators do not phase lock with each other, and the solution instead corresponds to a limit cycle in which only oscillators that are not directly coupled synchronize with each other (the groups $\theta_n=\theta$ and $\phi_n=\phi$ drift as in the subcritically coupled Kuramoto model with $N=2$). We refer to this state as the remotely synchronized limit cycle, following the nomenclature from the study of remote synchronization\cite{bergner2012remote}.  For $\sigma>0.25$, on the other hand,  phase locking between the groups results in stable and unstable synchronized steady-state branches. The stable and unstable synchronized states are mapped to each other under the time-parity reversal for $\sigma>0.25$, while they become invariant under the time-parity reversal for $\sigma=0.25$. The remotely synchronized limit cycle remains invariant for $\sigma<0.25$ and is neutrally stable at the phase-locking bifurcation point, with $2N-2$ nontrivial unity Floquet multipliers. We can observe numerically that the remotely synchronized invariant chimera retains all $2N-2$ unit Floquet multipliers below the SBP and remains neutrally stable. The SBP is therefore a saddle-node bifurcation on the invariant circle with codimension $N-1$, and many unstable solution branches emerge out of it which break the rotational and time-reversal symmetries. 

In addition to the synchronized solutions, twisted solutions with $\theta_n=\theta+2\pi p/N$ and $\phi_n=\phi+2\pi p/N$ for integers $p$, with analogous SBPs occurring for $\sigma>0.25$. Further cluster-twisted steady-state solutions can also be found, defined by the reduction $\theta_{n+q}=\theta_n+\nu q$ and  $\phi_{n+q}=\phi_n+\nu q$ for twisting parameter $\nu$ and number of clusters $N/q$ \cite{nicolaou2019multifaceted}. These twisted and cluster-twisted solutions also exhibit SBPs corresponding to synchronization transitions for $\sigma$ slightly above $\sigma=0.25$. No attracting steady-state or limit-cycle solutions are known for $\sigma<0.25$, and direct numerical simulations suggest the existence of a chaotic attractor that exhibits behavior qualitatively reminiscent of spatiotemporal intermittency in the complex Ginzburg-Landau equation\cite{chate1994spatiotemporal} near the synchronization transition.

In summary, the rotational and time-reversal symmetries in the ring of Janus oscillators lead to a new mechanism for the formation of localized oscillatory states. Figure \ref{fig2}(g) enumerates the key bifurcations of the various steady-state and limit-cycle solutions for the exemplary solution branch using the complex Kuramoto order parameter projection for visualization. Two families (parameterized by global phase rotations) of symmetry-related unstable fixed points emerge out of an SBP corresponding to the phase-locking synchronization transition at $\sigma=\sigma_{SBP1}=0.25$.  At slightly larger $\sigma=\sigma_{SBP2}\approx0.2620$, the period-tripled invariant limit-cycle solution undergoes another SBP, splitting into two symmetry-related limit cycles. By $\sigma=0.3$, one of the two limit cycles has become stable after undergoing several torus bifurcations. Finally, for yet larger $\sigma=\sigma_{HC}\approx0.3480$,  the limit-cycle solutions collide with their respective steady-state solutions in a heteroclinic bifurcation, forming two heteroclinic cycles. 

\subsubsection{Other solution branches}
While other solution branches are formed via mechanisms that are qualitatively similar to the exemplary branch, they also exhibit more complexity.  For example, for the top attracting chimera states, the heteroclinic cycle visits a greater variety of steady-state solutions.  This chimera state alternatingly visits distinct steady states which are related by a reflection before proceeding to visit the rotated states, as shown in Fig.~\ref{janus_invariants}(a).  Other branches are observed to visit multiple unstable steady states that are unrelated by symmetries, as in Fig.~\ref{janus_invariants}(b).

Figure \ref{janus_invariants}(c-e) shows three examples of the invariant chimera states that emerged out of SBPs. The traveling wave solution following the second SBP of the exemplary solution branch is shown in Fig.~\ref{janus_invariants}(c), which is invariant under $\pi_4$. In Fig.~\ref{janus_invariants}(d), the oscillators form two clusters of $8$ which differ by a phase shift of $\pi$, thus forming a limit-cycle solution in one of the cluster twisted reductions. with $q=8$ and $\nu=\pi/2$. Correspondingly, the Kuramoto order parameter is exactly zero for this invariant chimera state, and it corresponds to the orange line with $r=0$ in Fig.~\ref{fig2}(a). This state is invariant under the reversals $\pi_1$, $\pi_4$ and $\pi_5$. Several stable chimera states branch off of this cluster twisted invariant chimera state. We also highlight a second interesting traveling wave invariant chimera state in Fig.~\ref{janus_invariants}(e), which corresponds to the blue invariant chimera branch that is prominent below $\sigma=0.25$ in  Fig.~\ref{fig2}(a). This solution branch is also invariant only under $\pi_4$.

\section{Parametrically driven pendulum array}\label{sec:Pendulum}
\subsection{Previous observations}
Recent studies on heterogeneity-stabilized homogeneous states \cite{nicolaou2021heterogeneity} and anharmonic classic time crystals \cite{nicolaou2021anharmonic} employed a model describing a parametrically-driven array of pendula with alternating lengths. In this model, the angles $\theta_n$ and $\phi_n$ of the $n$th long and short pendula, respectively, evolve according to
\begin{align}
M \ddot{\theta}_n &= -\eta \dot{\theta}_n- \frac{Mg + 4\kappa\Delta +  Ma_d \omega_d^2 \cos(\omega_d t)}{1+\Delta}\sin(\theta_n) \nonumber \\
&\quad +\kappa \frac{1-\Delta}{1+\Delta} \left[\sin(\phi_n-\theta_n) + \sin(\phi_{n-1}-\theta_n)\right],  \label{pendula1} \\
M \ddot{\phi}_n &= -\eta \dot{\phi}_n- \frac{Mg - 4\kappa\Delta +  Ma_d \omega_d^2 \cos(\omega_d t)}{1-\Delta}\sin(\phi_n) \nonumber \\
&\quad +\kappa \frac{1+\Delta}{1-\Delta} \left[\sin(\theta_n-\phi_n) + \sin(\theta_{n+1}-\phi_n)\right], \label{pendula2}
\end{align}
where $\Delta$ is the alternating length scale, $a_d$ is the driving amplitude, and $\omega_d$ is the driving frequency. We fix the damping coefficient to $\eta=0.1$, the gravitational acceleration to $g=1$, the mass to $M=1$, and the coupling spring constant to $\kappa=1$ throughout these numerical investigations. For concreteness, we assume there are a finite number of $N \geq 1$ pairs of pendula, and we employ periodic boundary conditions with $n=N+m$ identified with $n=m$. 

The pendulum array exhibits relatively fewer discrete symmetries than the ring of Janus oscillators. Still, it does posses symmetries, as it is invariant under a reflection $(\theta_n',\phi_n') = \pi_1(\theta_n,\phi_n) \equiv (\theta_{N-n},\phi_{N-n-1})$, a second reflection $(\theta_n',\phi_n') = \pi_2(\theta_n,\phi_n) \equiv (-\theta_{n},-\phi_{n})$, and a translation symmetry $\pi_R$ given by $(\theta_n',\phi_n')=\pi_R(\theta_n,\phi_n)\equiv(\theta_{n+1},\phi_{n+1})$.  
\begin{figure}[t]
\includegraphics[width=0.9\columnwidth]{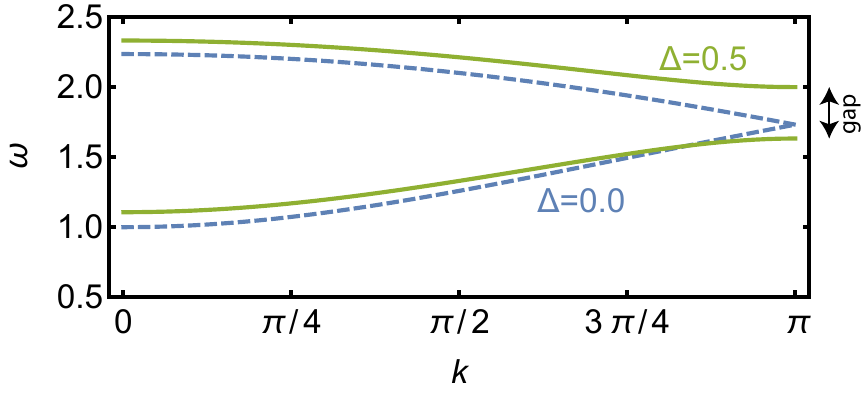}
\caption{Wave frequency $\omega$ vs wavenumber $k$ (the dispersion relation) for a homogeneous system ($\Delta=0$) and a heterogeneous system ($\Delta=0.5$) in the absence of driving ($a_d=0$). A band gap emerges in the heterogeneous case. \label{fig_bandgap}}
\end{figure}

The traditional study of parametric instabilities begins by examining the stability of the homogeneous state $\theta_n=\phi_n=0$. Taking advantage of the translational symmetry, we employ the Floquet wave mode ansatz
\begin{align}
\theta_n &= e^{\mathrm{i} kn+s t}\sum_m {\Phi} _{0 m} e^{ \mathrm{i} m \omega_d t}, \\
\phi_n &= e^{\mathrm{i} kn+s t}\sum_m {\Phi} _{1 m} e^{ \mathrm{i} m \omega_d t}.
\end{align}
Wave modes are characterized by the relationship between the wavenumber $k$, the frequency $\omega$, and growth rate $b$, where $s =  b + \mathrm{i}\omega$ is the complex Floquet exponent, related to the Floquet multiplier $\mu$ via $\mu=e^{s}$.  
The linearized equations 
\begin{equation}
\label{eigen2}
\sum_{i m} D^{i m}_{j n} {\Phi} _{i m} = a_d\sum_{i m} E^{i m}_{j n} {\Phi} _{i m}, 
\end{equation}
determine the stability of the homogeneous state, with
\begin{align}
D^{i m}_{j n} &= {L_i}\left[ M(s+{\mathrm{i}}\omega_d n)^2+\eta(s+{\mathrm{i}}\omega_dn)+2\kappa \right]\delta^i_j\delta^m_n \nonumber \\
 &\quad -{2\kappa L_i\left(\delta^i_{j+1} + \delta^i_{j-1}\right)}\cos (k) \delta^m_n + M g\delta^i_j\delta^m_n , \\
E^{i m}_{j n} &= {-}\frac{1}{2}M\omega_d^2\left(\delta^m_{n+1} + \delta^m_{n-1}\right)\delta^i_j. 
\end{align}
and $L_i=1+(-1)^i\Delta$. Solving Eq.~\eqref{eigen2} as a nonlinear eigenvalue problem for $s$ given $a_d$ gives a generalized dispersion relationship for the pendulum array.  For $a_d=0$, the growth rate for all modes is given by $-\eta/2$, so the homogeneous state is stable. As $a_d$ increases, instabilities occur when the growth rate for some mode becomes positive, leading to pattern-forming dynamics. The local bifurcations that give rise to the initial instability govern the kind of pattern formation observed in the pendulum array.
\begin{figure*}[htb]
\includegraphics[width=1.9\columnwidth]{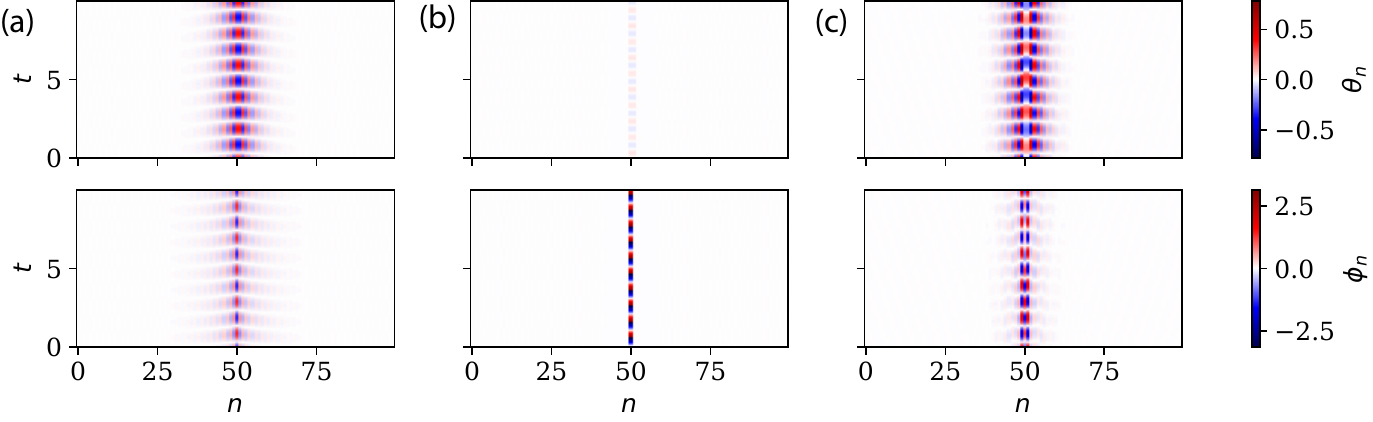}
\caption{Localized states in a parametrically-driven array of $N=100$ pendulum pairs. (a) Stable subharmonic localized states (which oscillate with half the driving frequency) observed from random initial conditions with $\Delta=0.25$ and $\omega_d=3.5$. (b) Stable harmonic localized state (which oscillates with the driving frequency) observed from random initial conditions with $\Delta=0.5$ and $\omega_d=3.35$.  (c)  Stable anharmonic localized state (which oscillates quasiperiodically with the driving frequency and an incommensurate second frequency) observed from random initial conditions with $\Delta=0.5$ and $\omega_d=3.35$.  (Multimedia view) Animation of the evolving phases for each solution.  \label{fig_pendula}}
\end{figure*}

The system symmetry, and hence the dynamics in the array, dramatically differs for $\Delta=0$ and $\Delta>0$. When $\Delta=0$, there is enhanced translational symmetry $(\theta_n',\phi_n')=(\phi_{n},\theta_{n+1})$ of half a unit cell.  Correspondingly, when $\Delta>0$, the wave modes split into two branches separated by a band gap in the wave frequency, as shown in Fig.~\ref{fig_bandgap}.  For most driving frequencies, the instability of the homogeneous state occurs as a wave mode with frequency $\omega=\omega_d/2$ resonates with the driving frequency. The growth rate $b$ of resonant modes quickly increases with increasing driving amplitude, giving rise to a period-doubling bifurcation when $b$ becomes positive. This instability mechanism is the same as the subharmonic response observed in the classical Faraday wave systems \cite{cross1993pattern}. However, when driving the system with a frequency corresponding to twice the band-gap frequencies, there are no wave modes that can easily resonate with the driving, and the band-gap opening gives rise to heterogeneity-stabilized homogeneous states\cite{nicolaou2021heterogeneity}.

When the heterogeneity-stabilized homogeneous states are perturbed by finite-amplitude disturbances, it is possible to excite a variety of oscillatory gap-soliton states in which the swinging amplitude is localized.  The gap solitons differ from chimera states because they generally do not exhibit coexisting synchronous and asynchronous domains, but they do exhibit localization. Three example solutions are illustrated in Fig.~\ref{fig_pendula}(Multimedia view), along with animations of the phases for each solution.   Figure \ref{fig_pendula}(a) shows a time-periodic state for $N=100$, $\Delta=0.25$, $\omega_d=3.5$, and $a_d=0.035$ which we found with random initial conditions for the pendulum angles drawn from a normal distribution with mean zero and standard deviation $0.5$.  The state is subharmonic, oscillating once every two driving periods,  and thus forms a localized limit cycle, with swinging amplitude quickly decaying to zero after about ten pendulum pairs in each direction.  For sufficiently large $\Delta$, on the other hand, it was also previously noted that responses that are qualitatively different from the classical subharmonic response can occur\cite{nicolaou2021anharmonic}. For example, Fig.\ref{fig_pendula}(b) and Fig.\ref{fig_pendula}(c) illustrate two cases with $\Delta=0.5$, $\omega_d=3.35$ and $a_d=0.05$ with the same kind of initial conditions as Fig.\ref{fig_pendula}(a). In Fig.~\ref{fig_pendula}(b), the central short pendulum rotates completely around its pivots once per driving cycle, exhibiting a localized state with a harmonic response to the driving. We also note the quasiperiod solution in Fig.~\ref{fig_pendula}(c), which we can regard as a localized anharmonic response.   Anharmonic wave mode responses to periodic driving in the pendulum array were recently characterized as emerging via a Neimark-Sacker bifurcation when two wave modes with the same wavenumber have a frequency difference equal to the driving frequency\cite{nicolaou2021anharmonic}, but their localized counterparts were not previously noted.   

\subsection{Continuation equations}
We again regularize with a complex representation $z_n=e^{\mathrm{i}\theta_n}$ and $w_n=e^{\mathrm{i}\phi_n}$ for numerical continuation. Since the model is second order in time, we also introduce the auxiliary momenta variables $p_n = (1+\Delta)\dot{\theta}_n$ and $q_n=(1-\Delta)\dot{\phi}_n$ to derive first-order equations.  Lastly, we introduce an auxiliary complex variable $Z$ evolving according to the Stuart-Landau equation, which will act as the periodic drive on the pendula. We then consider the complex equations of motion
\begin{align}
\dot{z}_n&=\mathrm{i}z_np_n/(1+\Delta)+\gamma(1-|z_n|^2)z_n \label{cpendula1} \\
M\omega_d\dot{p}_n&=-\eta p_n - (Mg+Ma_d\omega_d^2\frac{Z+Z^*}{2}+4\kappa\Delta)\frac{z_n-z_n^*}{2\mathrm{i}} \nonumber \\
&\quad +\kappa(1-\Delta)\frac{(w_n+w_{n+1})z_n^*-(w_n^*+w_{n+1})z_n}{2\mathrm{i}} \label{cpendula2} \\
\dot{w}_n&=\mathrm{i}w_nq_n/(1-\Delta)+\gamma(1-|w_n|^2)w_n \label{cpendula3}  \\
M\omega_d\dot{q}_n&=-\eta q_n-(Mg+Ma_d\omega_d^2\frac{Z+Z^*}{2}-4\kappa\Delta)\frac{w_n-w_n^*}{2\mathrm{i}} \nonumber \\
&\quad +\kappa(1+\Delta)\frac{(z_n+z_{n-1})w_n^*-(z_n^*+z_{n-1})w_n}{2\mathrm{i}} \label{cpendula4}\\
\dot{Z}&=\mathrm{i}Z+\gamma(1-|Z|^2)Z. \label{cpendula5}
\end{align}

The driving variable $Z$ is decoupled from the other equations and quickly relaxes to the limit-cycle attractor $Z=e^{\mathrm{i}\tau}$, where $\tau\equiv \omega_d t$ is the non-dimensional time. Thus, the terms $(Z+Z^*)/2$ in Eqs.~\eqref{cpendula2} and \eqref{cpendula4} reduce to $\cos (\omega_d t)$, which acts as the parametric driving term, and the phase equations reduce to a non-dimensionalized version of the pendulum array Eqs.~\eqref{pendula1}-\eqref{pendula2}. We express Eqs.~\eqref{cpendula1}-\eqref{cpendula5} in Cartesian coordinates to continue the resulting system of $6N+2$ real equations in AUTO. Numerical continuation of invariant tori like the novel state in Fig.~\ref{fig_pendula}(b) is not yet implemented in AUTO and is very costly to perform. Thus, we focus in the remainder of the paper on limit cycle solutions only.
\begin{figure*}[t!]
\includegraphics[width=1.5\columnwidth]{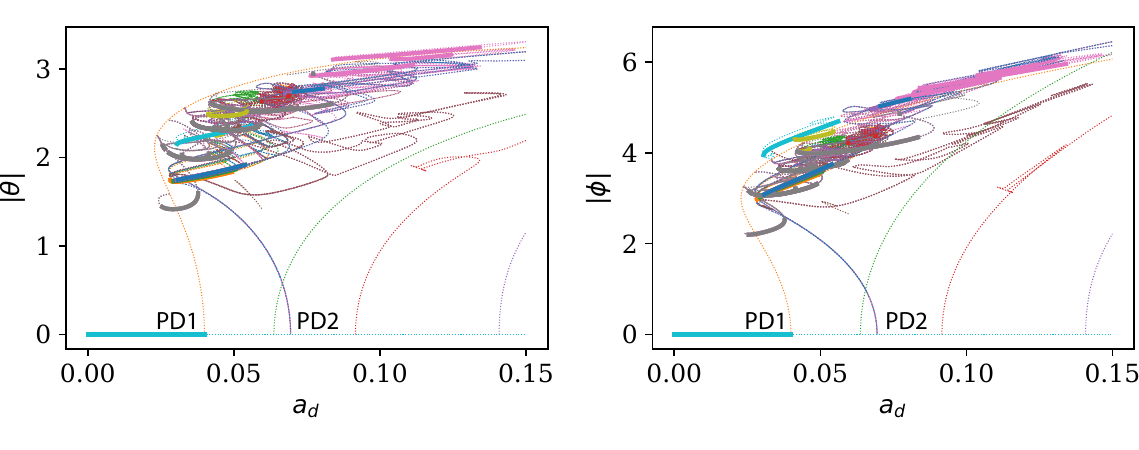}
\caption{Bifurcation diagram for the parametrically-driven pendulum array with $\omega_d=3.5$ and $\Delta=0.25$. Thick lines show stable limit cycles, thin dotted lines show unstable limit cycles, and the primary subharmonic period-doubling bifurcations PD1 and PD2 preceding localization are marked. \label{fig3}}
\end{figure*}

\subsection{Continuation results}
For simplicity, we fix $\Delta=0.25$  and $\omega_d=3.5$ (with $\omega_d/2$ lying within the band gap) while continuing solutions in the driving amplitude $a_d$.  Because numerical continuation for limit cycles is expensive, we must use a relatively small value of $N=16$ here,  but we suggest that the localized states like that in Fig.~\ref{fig_pendula}(a) and their related bifurcations will not significantly differ from their large $N$ counterparts. We begin again by identifying stable periodic orbits at $a_d = 0.045$ from evolving $10000$ random initial conditions over a period of $5000$ driving periods. We bin the final states according to the sorted and time-averaged squared angles of the pendula and identify the $25$ top attracting initial limit cycles to continue. Stable (unstable) solutions are shown as thick (thin) lines in Fig.~\ref{fig3}, where the solution norms are $|\theta| = \left(\int \sum_n \theta_n(t)^2 dt\right)^{1/2}$ and $|\phi| = \left(\int \sum_n \phi_n(t)^2 dt\right)^{1/2}$.

The homogeneous state corresponds to solutions with $|\theta|=|\phi|=0$. It has a period equal to the driving period (only the auxiliary variable $Z$ varies over the period). This homogeneous state first becomes unstable in a period-doubling bifurcation, labeled PD1 in Fig.~\ref{fig3}(a) and occurring at $a_d\approx0.0399$. All other limit cycle solutions in Fig.~\ref{fig3} have a period twice the driving period. The initial period-doubling bifurcation is subcritical (since $\omega_d/2$ lies in the band gap) and corresponds to a spatial wavenumber $k=\pi$, with an unstable swinging branch emerging for lower driving amplitudes (orange dotted line).  Various solution branches emerge off this period-doubled $k=\pi$ branch in secondary bifurcations. A second subcritical period-doubling bifurcation labeled PD2 in Fig.~\ref{fig3}(a) and occurring at $a_d\approx0.0694$ corresponds to the wavenumber $k=7\pi/8$ and undergoes similar bifurcations of its own. The secondary solution branches following these bifurcations are interconnected with the previous ones in a complicated tangle of solutions and bifurcations.

\begin{figure}[h!]
\includegraphics[width=0.75\columnwidth]{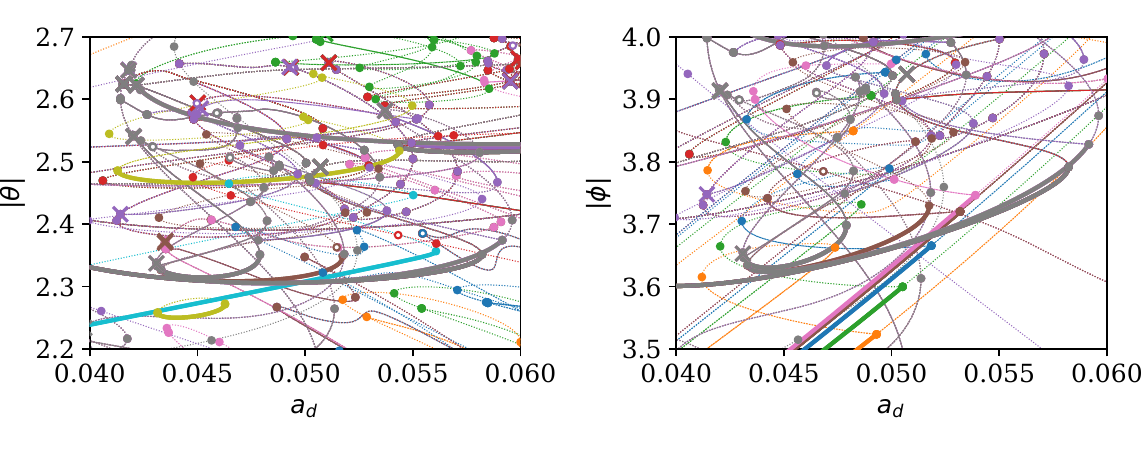}
\caption{Zoomed-in portion of the bifurcation diagram in Fig.~\ref{fig3}, with fold and simple branch bifurcation points marked by solid circles, torus bifurcation points marked by open circles, and SBPs marked by the $\times$ symbols.  \label{figtangles}}
\end{figure}

Figure.~\ref{figtangles} shows a zoomed-in portion of all stable and unstable solution branches.  Our calculations were restricted to twenty branch points and twenty limit points before the continuation was terminated, but it appears that an increasingly large number of branch points would emerge with increasing resolution. These tangles resemble the homoclinic tangles seen in chaotic systems such as the Smale horseshoe, and we conjecture that the full set of solution branches in the bifurcation diagram forms a fractal set.
\onecolumngrid

\begin{figure}[t!]
\includegraphics[width=0.75\columnwidth]{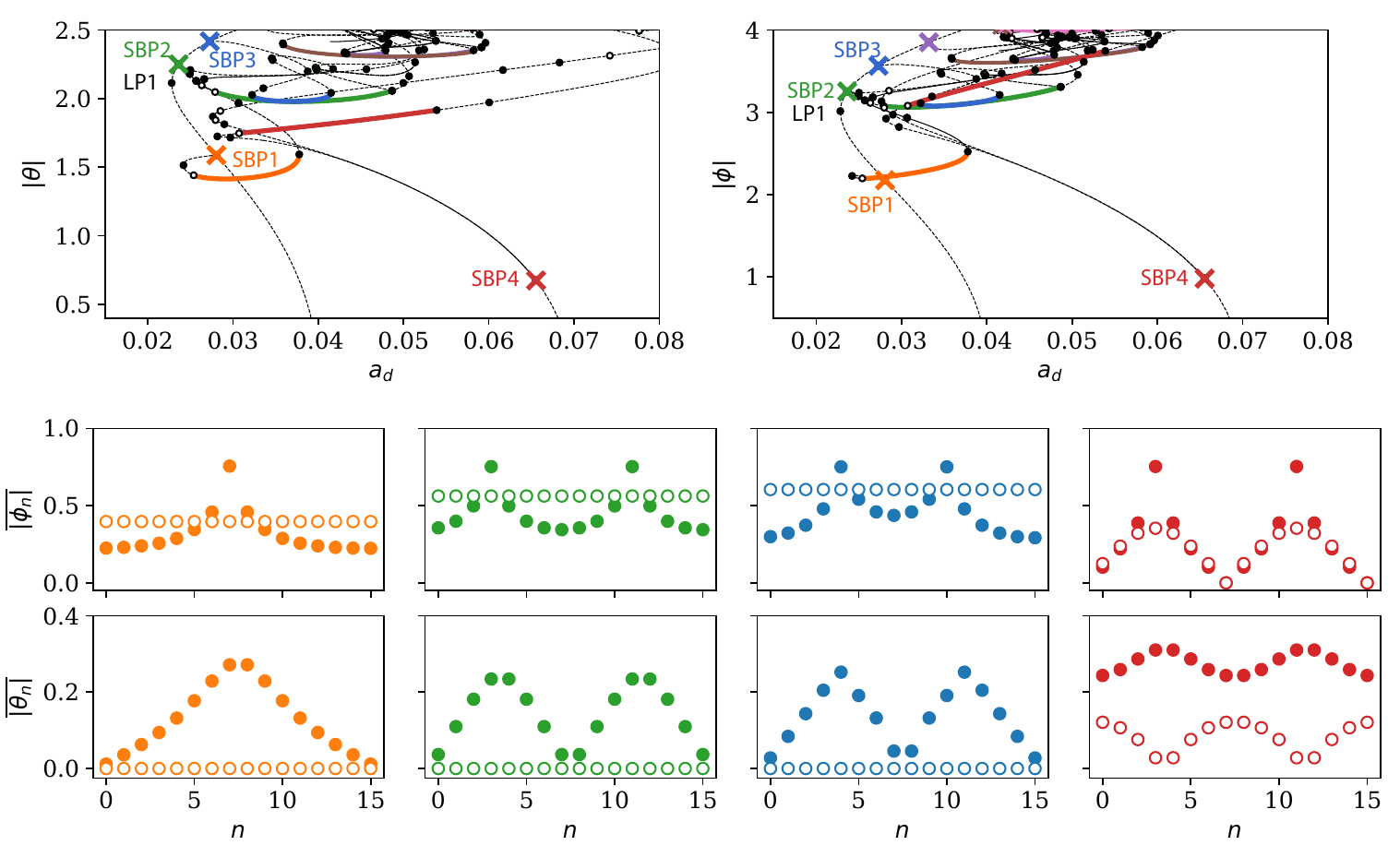}
\caption{Examples of localized states emerging from SBPs in the pendulum array. The top row shows a blow-up of the bifurcation diagrams with only pertinent branches shown,  as in Fig.~\ref{figtangles}. The lower two rows show the time-averaged swinging amplitudes for the color-corresponding SBPs (open circles) and a stable solution (filled circled) along the following stable solution branches at $a_d=0.037$.  \label{figexamples}}
\end{figure}
\twocolumngrid

Figure \ref{figexamples} shows a few example solution branches in greater detail (with extraneous unstable branches omitted for clarity). The $k=\pi$ period-doubled branch exhibits a limit point labeled LP1 but does not become stable when it turns around here. This is because an additional branching bifurcation labeled SBP1 in Fig.~\ref{figexamples} occurs at $a_d\approx 0.02806$ on the solution branch before the limit point. Similar secondary branching off of a subcritical solution branch has been reported to prevent periodic stabilization in binary fluid convection\cite{batiste2001simulations}.  The unstable symmetry-broken solution branch that emerges from the SBP1 becomes stable after turning around and undergoing a torus bifurcation.  The color-corresponding lower panels of Fig.~\ref{figexamples} show the solution at SBP1 (open circles) and the solution after stabilizing (closed circles). This stable solution branch corresponds to a localized state centered around a single group of swinging pendula. Since our continuations are expensive, we restrict the number of pendula pairs to $N=16$ here, but, in larger arrays, the state is indeed localized in the sense that swinging amplitudes decay towards zero as one moves away from the localization center, as in Fig.~\ref{fig_pendula}(a).  Along the $k=\pi$ branch, the following SBP2 and SBP3 bifurcations occur at $a_d\approx 0.02363$ and $a_d\approx 0.02729$, respectively,  and their corresponding stabilized solutions correspond to localized states with two localization centers with differing symmetry. The point SBP4 occurring at $a_d=0.06944$ and the corresponding stabilized solutions emerging from the $k=7\pi/8$ branch have similar symmetry to those corresponding to SBP2 but with differing swinging amplitudes. Further SBPs occur farther along the solution branches and correspond to solutions with a larger number of localization centers.

In summary, localized states emerge in the pendulum array following SBPs on the subcritical branches of period-doubled wave modes. These interconnected solution branches bear some resemblance to the snaking bifurcations in the Swift-Hohenberg equation, but they are not nearly as organized. Furthermore,  a very large number of unstable periodic solutions branch out of the stable localized states and form a tangle of unstable solutions, resembling the homoclinic tangles in chaotic systems.

\section{Strange snaking bifurcations}\label{sec:BifDiags}
Here we attempt to rationalize the complex bifurcation observed in the ring of Janus oscillators and the array of coupled pendula. For illustration, we return to the pseudoarclength continuation method for steady-state solutions described in Sec.~\ref{continuationsec} for Eq.~\eqref{CoupODE}. We note that while continuation is typically implemented as an iterative algorithm for solving an extended system, we can consider the process as a dynamical system in its own right. 
Consider, for example, the dynamical system corresponding to the standard pseudoarclength continuation technique
\begin{align}
 \begin{pmatrix} J_{nm} && \partial F_n/\partial \mu + \varepsilon\sum_k \partial G_n / \partial \mu \\ \delta x_m && \delta s \end{pmatrix} \begin{pmatrix} {\partial x_m}/{\partial s} \\ {\partial \mu}/{\partial s} \end{pmatrix}= \begin{pmatrix} 0 \\ 1 \end{pmatrix}. \label{pasys}
\end{align}
Here, the direction vector $\begin{pmatrix} \delta x_m & \delta s \end{pmatrix}^\top$  is the normalized (right) null vector of the matrix $N\times (N+1)$ submatrix $\begin{pmatrix} J_{nm} && \partial F_n/\partial \mu + \varepsilon \sum_k \partial G_n / \partial \mu \end{pmatrix}$, which is guaranteed to be unique up to sign at regular solution points and simple branch points \cite{doedel1991numerical}. The first $N$ rows in Eq.~\eqref{pasys} ensure that the values of $F+\varepsilon \sum_k G$ do not vary along the trajectory, and the last row ensures a constant rate of (pseudo)arclength increase. In generic systems (with only regular solution points potentially exhibiting fold and Hopf bifurcations), it is always possible to invert the matrix in Eq.~\eqref{pasys} at the fixed points, and so the system is well defined at least in a neighborhood of all the fixed points. 

When viewed as a nonlinear dynamical system, it is easy to see that the trajectories defined by Eq.~\eqref{pasys} can be quite complicated. For generic systems, this dynamical system can exhibit no fixed points since the inverse matrix cannot have zero eigenvalues, which constrains the trajectories significantly. But for symmetric systems, the extended Jacobian need not be invertible, and the dynamics can be complex, perhaps even approaching a chaotic attractor as $s$ increases.  The bifurcation diagram for Eq.~\eqref{CoupODE} would then be \textit{strange}, exhibiting snaking branches that entwine endlessly. In such a scenario, the original dynamics in Eq.~\eqref{CoupODE} would necessarily possess infinitely many fixed points for some values of $\mu$, but this can be possible even if all the attractors are strictly fixed points. This phenomenon differs from spatial analogs of chaos, in that sensitive dependence arises in variations in parameter values rather than sensitive dependence on spatial location,  implying unpredictable switching dynamics between multistable states, for example.

We thus propose that nonattracting chaotic invariant sets (chaotic saddles) coexist with the stable limit cycle solutions in  Eq.~\eqref{CoupODE} in our case studies.   Such chaotic sets are typically multifractal and possess a skeleton of embedded unstable periodic orbits that define their geometry \cite{ott2002chaos}. Chaotic saddles undergo sudden transitions (crises) when their unstable periodic orbits interact with other, external invariant sets.  Thus, we suggest that the stable periodic orbits observed in our models lose their stability by interacting with unstable periodic orbits that go on the be involved in crises with a chaotic saddle. In this case, numerical continuation starting from stable states would eventually lead to the skeleton of the chaotic saddle and could exhibit strange bifurcation diagrams as we observe.

\section{Discussion}\label{sec:Discussion} 
In this paper, we documented novel mechanisms for localized pattern formation in systems of coupled oscillators. In particular, we have shown that the emergence and bifurcation structure of the multitude of stable localized states herein is significantly different from the well-documented steady states in other pattern-forming systems. We produced nontrivial continuations of periodic and traveling states that required modifications to existing numerical software, highlighted connections with heteroclinic networks and nonattracting chaotic invariant sets, and provided some rationale for the strange bifurcation diagrams that we observed.  We suggest that these case studies may help shine a light on the complex multistable switching dynamics mediated by chaos in other systems that have recently attracted interest\cite{ansmann2016self,bick2018heteroclinic,zhang2020critical,medeiros2023chimera}.

Our studies revealed several specific areas for potential improvement in numerical continuation with AUTO. First, we found that detecting simple branch points via the determinant of the extended system Jacobian is a limitation, both because of poor numerical stability and because of the inability to detect SBPs, where multiple eigenvalues or Floquet multipliers simultaneously change stability. It would be desirable to instead efficiently sort the eigenvalues and detect points where individual or groups of eigenvalues change sign, which would also help to distinguish the local structure and symmetry properties of the bifurcations.  A second challenge arises in systems with time-reversal symmetries, like PT symmetry of interest in topological matter systems. We found in the Janus ring that invariant solution branches can have several neutrally stable Floquet multipliers which are constrained to the unit circle by the symmetry. To detect bifurcations in such solutions, it would be desirable to automatically track the number of symmetry-constrained neutral multipliers as well so that their spurious sign changes can be ignored in the bifurcation detection. Preliminary efforts to implement each of these improvements are described in our GitHub repository \cite{github}, which we anticipate may aid in the study of other systems or in a follow-up investigation that provides a detailed and exhaustive study of the many parameter regimes for our Janus ring and pendulum models that were not explored here.

Beyond follow-up numerical studies, what remains is a full analytical investigation that can explain our observations in some level of generality. That is, it is important to know the classes of systems where one should expect to observe certain localized patterns and understand all of the mechanisms that lead to their formation. For example, it would be very valuable to characterize the details of the heteroclinic bifurcations\cite{champneys2009unfolding} involved in the formation of the localized traveling chimeras in the ring of Janus oscillators. Moreover, we would like to know why some patterns travel, as in the Janus ring, while regions of localization can also remain fixed at certain indices, as in the pendulum. We anticipate a partial explanation for the traveling could come from recent work on the Swift--Hohenberg equation where it was shown that breaking the variational structure of the equation leads to traveling asymmetric states \cite{raja2023collisions}. Similarly, decades of analytical advancement in the understanding of localized steady-state formation can be used to inform and contrast with results on localized oscillations in systems of coupled oscillators. Finally, traveling localized defects such as those documented throughout this paper are common in partial differential equations supporting waves \cite{sandstede2004defects}. So,  it may be the case that in certain parameter regimes, our oscillator systems can be thought of as discretizations of spatially extended systems. Such a connection would allow for the application of well-developed analytical techniques from partial differential equations to be brought over to the systems herein, but it remains to be established.   

\begin{acknowledgments}
ZGN is a Washington Research Foundation Postdoctoral Fellow. JJB is supported in part by an NSERC Discovery Grant.
\end{acknowledgments}

\section*{Data Availability Statement}
All data in this paper can be reproduced from the source code in our GitHub repository \cite{github}.

\bibliography{oscillatorSnaking.bib}

\end{document}